\documentclass[aps,pre,twocolumn]{revtex4-1}

\usepackage{graphicx}
\usepackage{amssymb}
\usepackage{amsmath}
\usepackage{verbatim}

\newcommand{\nn}{\nonumber}
\newcommand{\veps}{\varepsilon}
\newcommand{\vphi}{\varphi}

\renewcommand{\eqref}[1]{Eq. (\ref{#1})}
\newcommand{\figref}[1]{Fig. \ref{#1}}

\begin{document}

\title{Stabilization of charged and neutral colloids in salty mixtures}

\author{Sela Samin}
\author{Yoav Tsori}
\email{tsori@bgu.ac.il}

\affiliation{Department of Chemical Engineering and The 
Ilse Katz Institute for Nanoscale Science and Technology, Ben-Gurion University of the 
Negev, 84105 Beer-Sheva, Israel.}

\date{\today}

\begin{abstract}

We present a mechanism for the stabilization of colloids in liquid mixtures without use of
surfactants or polymers. When a suitable salt is added to a solvent mixture, the coupling
of the colloid's surface chemistry and the preferential solvation of ions leads to a
repulsive force between colloids that can overcome van der Waals attraction. This
repulsive force is substantial in a large range of temperatures, mixture composition and
salt concentrations. The increased repulsion due to addition of salt occurs even for
charged colloids. This mechanism may be useful in experimental situations where steric
stabilization with surfactants or polymers is undesired. 

\end{abstract}

\maketitle

The stability of colloidal dispersions is essential in material
science and technology. Steric stabilization of colloids 
against the attractive van der Waals (vdW) forces can be achieved using surfactant or
polymer molecules that are physically or chemically attached to the colloid's surface.
Charged colloids can also be stabilized via the screened Coulomb repulsion, whose range
depends on the Debye length $\kappa^{-1}$. In the celebrated Derjaguin, Landau,
Verwey, and Overbeek (DLVO) theory, addition of salt to the suspension decreases the 
Debye length and the electrostatic repulsion leading eventually to
coagulation and sedimentation of the colloids \cite{colloids_book}.

In recent years we began to better understand the differences 
between
the electrostatics of pure solvents compared to liquid mixtures. The preferential wetting
of one liquid component at the colloid surface \cite{Beysens1998,Evans2009} affects the
density of the ions, the electrostatics of the mixture, and the interaction between the
colloids 
\cite{Beysens1985,Beysens1991,leunissen2007,leunissen2007b,Bonn2009,Bechinger2008,
nellen2011,nguyen2013,shelke2013}. A
key parameter is the selective solvation of the ions in the
liquids 
\cite{onuki:3143,tsori_pnas_2007,zwanikken2007,zwanikken2008,efips_epl,onuki2011,onuki2011b,bier2011,
efips_jcp_2012,
bier2012}.

In this paper we present a new method for the stabilization of electrically charged or 
neutral particles in solvent mixtures by \emph{addition} of salt, without use of 
surfactants or polymers. The stabilization relies on (i) selective adsorption of one 
solvent on the colloids and (ii) a difference in the preferential solubilities of the 
anion and the cation in the solvents, namely a difference between the Gibbs transfer 
energy of moving the anion from one solvent to the other and the analogous Gibbs 
energy of the cation. These energies are often larger than the thermal energy 
\cite{marcus_cation,marcus_anion} and therefore particles can be stabilized even without 
other additives. In fact, requirements (i) and (ii) above are generally met because any 
surface is hydrophilic or hydrophobic to some extent and no two ions have identical 
solvation energies \cite{marcus_cation,marcus_anion}. The coupling of the colloid's 
surface chemistry and the preferential solvation of ions leads to a repulsive force 
between colloids that can overcome the van der Waals attraction. This repulsive force is large 
in a unexpectedly wide range of temperatures, mixture compositions and salt 
concentrations. This method preserves the chemical properties of the surface and therefore 
is advantageous in cases where surfactants, polymers, or a chemical modification of the 
particle are undesired.

We consider relatively dilute mixtures in which ion-ion correlations effects are small 
\cite{kanduc2012,cruz2013}. In order to isolate the effect of preferential solvation we 
ignore specific ion-surface interactions \cite{levin2011,markovich2013}. These short range 
interactions can either enhance or counteract the stabilization mechanism we discuss and 
their magnitude is of specific nature.

We focus on a system composed of a binary aqueous mixture
containing a $1:1$ monovalent
salt and confined between two identical flat plates. The
plates are located at $z=\pm D/2$ and their area is $S$. The mixture
composition is given by the water volume fraction $\phi$ ($0\le\phi\le1$)
while the cosolvent composition is given by $1-\phi$. The number
densities of the point-like positive and negative ions are denoted by $n^\pm$.
The fluid between the plates is in
contact with an electroneutral matter reservoir at composition $\phi_0$
and a salt concentration $n_0$.

The mean-field free energy density of the system is
given by \cite{efips_epl}:
\begin{align}
f&=k_BT \left[f_m(\phi)+\frac12 C|\nabla
\phi|^2\right]-\frac{1}{2}\veps_0\veps(\phi)(\nabla
\psi)^2 \nn \\ &+e(n^+-n^-)\psi+k_BT\sum_{i=\pm} n^i\left[\left(\log
(v_0n^i)-1\right)-\Delta
u^i\phi \right],
	\label{eq:fmions}
\end{align}
where $k_B$ is the Boltzmann constant, $T$ is the temperature, $\veps_0$ is the vacuum
permittivity, and $v_0=a^3$ is the molecular volume. $v_0f_m$ is the dimensionless mixing free energy density:
$v_0f_m=\phi\log(\phi)+ (1-\phi)\log(1-\phi)+\chi\phi(1-\phi)$, where the Flory parameter
is $\chi\sim1/T$. The mixture demixes for $T<T_c$ ($\chi>\chi_c=2$). The energetic cost of
composition inhomogeneities is accounted for by the square-gradient term, where $C$ is a
positive constant with units of inverse length. In the electrostatic energy in \eqref{eq:fmions}, $\psi$ is the
electric potential and $\veps$ is the dielectric constant, assumed to depend linearly on
composition by $\veps(\phi)=\veps_c+(\veps_w-\veps_c)\phi$, where $\veps_w$ and $\veps_c$
are the water and cosolvent dielectric constants, respectively. 
The first term on the second line of \eqref{eq:fmions} is the ions' 
electrostatic energy, where $e$ is the elementary charge. The first term in the sum 
is the ideal-gas entropy of the ions and the second term is  
the ion solvation energy. In our simple theory, the solvation 
energy is proportional to the local solvent
composition and its strength is measured by the parameters $\Delta u^i$
\cite{tsori_pnas_2007,onuki:3143}. Here we are interested in salts where 
the asymmetry in the cation and anion solvation
parameters, defined as $\upsilon \equiv (\Delta u^+-\Delta u^-)/2$, is large. This
is commonly satisfied in antagonistic salts where one ion is hydrophilic and the other
is hydrophobic: $\Delta u^+ \Delta u^- <0$.

The short range and electrostatic interactions between the fluid and the solid
surfaces are given by the surface free energy density $f_s:$
\begin{equation}
f_s=k_BT\Delta\gamma\phi({\bf r}_s)+e\sigma\psi({\bf r}_s),
\label{eq:fsions}
\end{equation}
where ${\bf r}_s$ is a vector on the colloid surface and
$e\sigma$ is the surface charge density of the plates. The surface wettability is given by
the parameter $\Delta\gamma$ that measures the difference between the solid-water and
solid-cosolvent surface tensions.                       
                       
The equilibrium state of the system is found by extremization of the grand potential
\begin{equation}
\Omega=\int\left[ f-k_BT(\lambda_0^+n^++\lambda_0^-n^-+\mu_0\phi) \right] {\rm d}
\mathbf{r} +\int f_s {\rm d}{\bf r}_s~,\nn
\end{equation}
where $\lambda_0^\pm$ and $\mu_0$ are the chemical
potentials imposed by the species in the reservoir. This leads to the first Euler-Lagrange
(EL) equation $\delta \Omega/\delta \phi=0$:
\begin{align}
\label{eq:comp}
-C
\nabla^2\phi+\frac{\partial f_m}{\partial \phi}
-\sum_{i=\pm}\Delta
u^in^i-\veps_0\frac{d\veps/d\phi}{2k_BT}(\nabla\psi)^2=\mu_0,
\end{align}
with the boundary condition $\mathbf{n} \cdot \nabla \phi = \Delta\gamma/C$, where ${\bf
n}$ is a unit vector perpendicular to the surface. The EL equation for the
potential $\delta \Omega/\delta \psi=0$ naturally yields Gauss' law: $-\nabla \cdot
(\veps_0\veps(\phi)\nabla \psi)=e(n^+-n^-)$ with the boundary condition
$-\mathbf{n} \cdot \nabla
\psi = e\sigma/\veps_0\veps(\phi)$. The densities $n^\pm$ from $\delta
\Omega/\delta n^\pm=0$
obey the Boltzmann distribution $n^\pm=v_0^{-1}\exp(\mp\Psi+\Delta
u^\pm\phi+\lambda_0^\pm)$, 
where $\Psi=e\psi/k_B T$ is the dimensionless potential. 

%
\begin{figure}[!tb]
\centering
\includegraphics[width=3.3inin,clip]{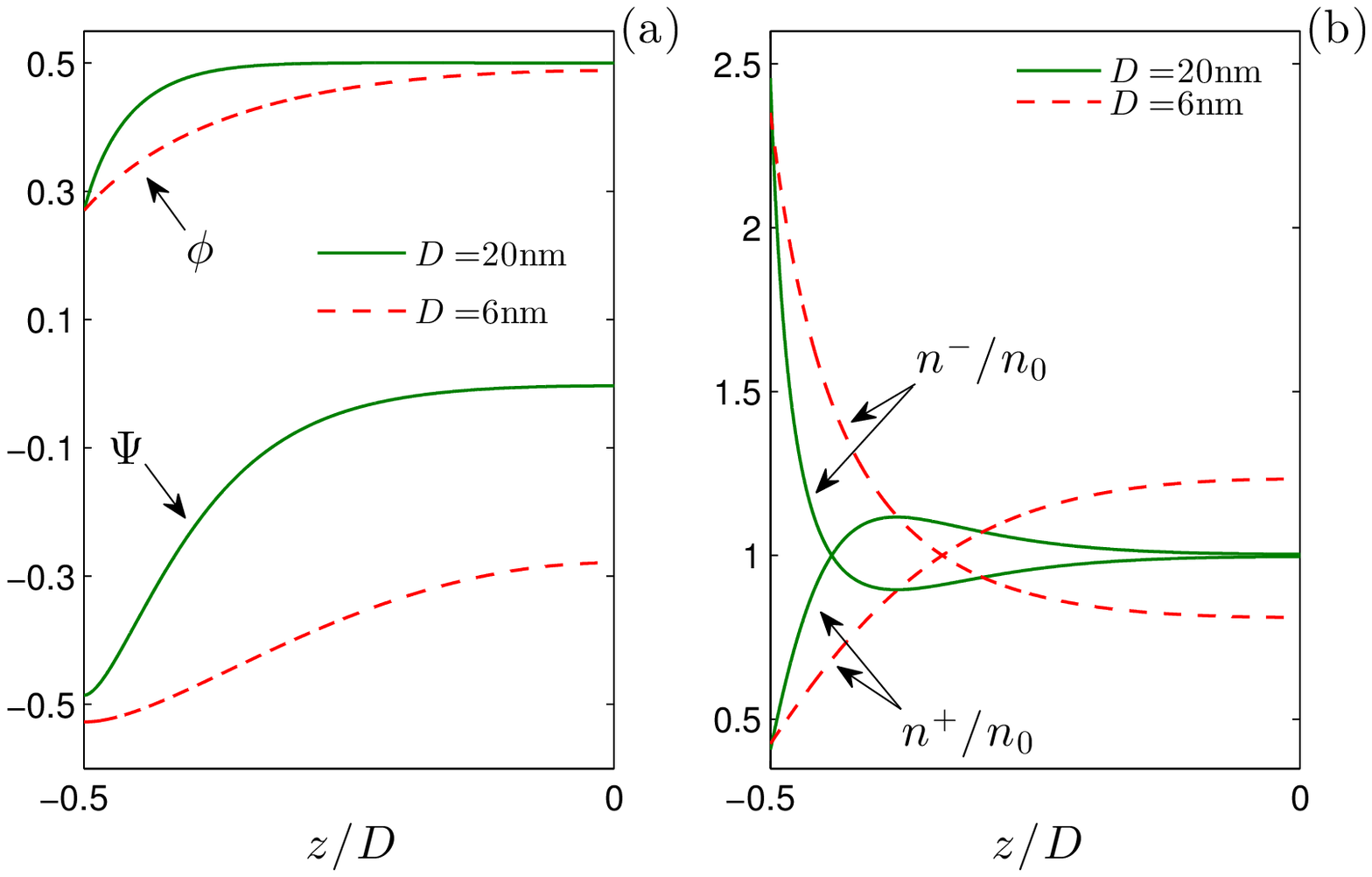}
\caption{Profiles of the (a)
composition, dimensionless potential and (b) scaled ion densities between two electrically
neutral plates immersed in a critical mixture ($\phi_0=1/2$) at a
temperature $\Delta T = T-T_c=21.5$K and containing $20$mM of antagonistic
salt ($\Delta u^+=-\Delta u^-=\upsilon=6$). Here and in other figures
the plates are hydrophobic with $\Delta\gamma=0.2/a^2$, corresponding to
about $7$ mN/m. As an approximation to water--2,6-lutidine mixtures we used
$T_c=307.2$K, $v_0=39$\AA{}$^3$, $C=\chi/a$ \cite{safran}, $\veps_{\rm
lutidine}=6.9$ and $\veps_{\rm
water}=79.5$.} 
\label{fig_stab1} 
\end{figure}

A solution of the EL equations in a planar geometry yields the one dimensional profiles
$\phi(z)$, $\psi(z)$ and $n^\pm(z)$. In \figref{fig_stab1} we plot the resulting profiles for a salty 
mixture between two electrically neutral and hydrophobic plates with a wettability $\Delta\gamma=0.2/a^2$. 
The mixture has a bulk critical composition, $\phi_0=1/2$, and its temperature is far above $T_c$, 
$\Delta T = T-T_c=21.5$K. The bulk concentration of the antagonistic salt is $n_0=20$mM 
and the solvation parameters are $\Delta u^+=-\Delta u^-=\upsilon=6$.

The ionic profiles in
\figref{fig_stab1} (b) show that an electrostatic diffuse layer (EDL) is realized near
the plates, despite their electric neutrality. The reason is the adsorption of the
cosolvent on the hydrophobic plates shown by the profile of $\phi$ in \figref{fig_stab1}
(a). The cosolvent ``drags'' the hydrophobic anions and repels the
hydrophilic cations. Hence, a net charge density develops in
the vicinity of the plates, giving rise to the electric potential profile shown in
\figref{fig_stab1} (a). 

The width of the adsorbed fluid layer is comparable to the bulk
correlation length $\xi$. Beyond this distance $\phi(z)$ decays
to its bulk value. However, electroneutrality dictates that the ionic profiles
must compensate for the deviation from the bulk values near the plate. At surface
separation of $D=20$nm (\figref{fig_stab1} (b), solid curves), this means that
the anion (cation) concentration becomes smaller (larger) than the bulk
value $n_0$ before it decays to $n_0$ at the midplane ($z=0$), leading to a
minimum (maximum) in the profile. On the other hand, at a
distance of $D=6$nm, the extremum is missing since the EDLs from each plate
overlap, such that $n^\pm(z=0) \neq n_0$. Such an overlap implies a repulsive
osmotic force between the plates.

Using the profiles we calculate the osmotic
pressure $\Pi$ between the plates at a given distance $D$ from 
$\Pi(D)=P_{zz}-P_0$, where $P_{zz}=\phi\delta f/\delta
\phi+n^+\delta f/\delta
n^++ n^- \delta f/\delta
n^--f-\veps_0\veps (\partial \psi/\partial z)^2$ is the $zz$ component of the Maxwell
pressure tensor
\cite{efips_epl} 
and $P_0=P_{zz}(\phi_0,n^{\pm}_0,\psi=0)$ is the bulk pressure. The interaction
potential $U(D)$ between the
plates is obtained from the osmotic pressure via $U(D)=-S\int_{\infty}^{D}
\Pi (D') \rm{d} D'~$.

\begin{figure}[!bt]
\centering
\includegraphics[width=3.3in,clip]{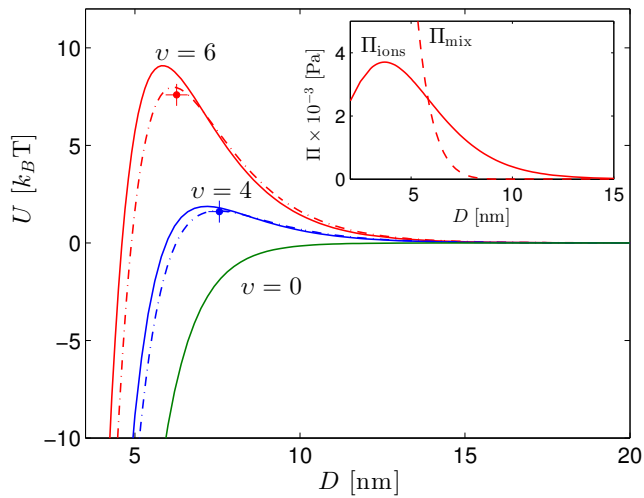}
\caption{(color online) The interaction potential $U(D)$ between electrically neutral and
hydrophobic plates immersed in a mixture with $\phi_0,T$ and $n_0$ as in
\figref{fig_stab1}. When both ions are hydrophilic the
interaction is purely attractive ($\Delta u^\pm=6$, green curve). 
A weak repulsive barrier $U_{\rm max}$ appears
with antagonistic salts ($\Delta u^+=-\Delta u^-=4$, blue curve), and its height
increases with increasing $\Delta u^+-\Delta u^-$ ($\Delta u^+=-\Delta u^-=6$, red
curve). Full numerical solutions (solid curves) are in good
agreement with linear theory (dash-dot curves), where we
used $\kappa^{-1}=1.68$nm, $\xi=0.94$nm and $l=1.78$. Two crosses are the
approximation
$(D_{\rm max},U_{\rm max})$ of \eqref{eq:analudmax}. Inset: The two components of the
osmotic pressure (see text) for the red curve. Here and in \figref{fig_stab3}
the area of the hydrophobic plates is
$S=0.01\mu$m$^2$ and $\Delta\gamma=0.2/a^2$.
} 
\label{fig_stab2} 
\end{figure}

The interaction potential for the parameters of \figref{fig_stab1} with solvation
asymmetry $\upsilon=6$ is plotted in \figref{fig_stab2} (red curve). At close separations
the potential is attractive due to adsorption of the cosolvent on the plates, but a
repulsive barrier of $\approx 10k_BT$ appears in $U(D)$ at a distance denoted by $D_{\rm
max}$ of a few nanometers. The barrier height $U_{\rm max}=U(D_{\rm max})$ strongly
depends on $\upsilon$. For an antagonistic salt with $\upsilon=4$ (blue curve) the barrier
is much smaller, while for a hydrophilic salt with $\upsilon=0$ (green curve), the
potential is purely attractive. 

In order to better understand the physical origin of the repulsion we
examine the components of the osmotic pressure. For symmetric plates, $\Pi$ can
be
recast in term of the mid-plane composition $\phi_m=\phi(z=0;D)$ and ion
densities $n^\pm_m=n^\pm(z=0;D)$ as 
$\Pi(D)=\Pi_{\rm ions}(n^\pm_m)-\Pi_{\rm mix}(\phi_m)$, where
\begin{align}
\Pi_{\rm ions}&=k_BT\left(n^+_m+n^-_m-2n_0\right), \\
\Pi_{\rm mix}&=k_BT\left(f_m(\phi_m)-f_m(\phi_0)-\mu_0(\phi_m-\phi_0)
\right).
\end{align}
The inset of \figref{fig_stab2} shows $\Pi_{\rm ions}$ and
$\Pi_{\rm mix}$ for the $\upsilon=6$ potential curve. It is seen that $\Pi_{\rm ions}$
is repulsive at $D\gtrsim4$nm; when the EDLs overlap this leads to an
increase of $n^\pm_m$. $-\Pi_{\rm mix}$ on the other hand is attractive. Since $U$
is the
cumulative integral of $\Pi_{\rm ions}-\Pi_{\rm mix}$, a repulsive barrier is
created by a range of $D$ for which $\Pi_{\rm ions}>\Pi_{\rm mix}$,
corresponding to $D\gtrsim6$nm in the Figure.

To better characterize the repulsive potential we solve the governing equations in the
limit where the perturbations in the composition $\vphi=\phi-\phi_0$ and ion densities
$\delta n^\pm = n^\pm-n_0$ are small \cite{onuki2011}: $\delta n^\pm=n_0(\Delta u^\pm
\vphi\mp\Psi)$, $\kappa^{-2}\nabla^2\Psi=\Psi-\upsilon \vphi$, $\kappa^{-2}
\nabla^2\vphi=l^{2}\vphi+ \omega^2(\Psi/\upsilon- \vphi)$, where $\kappa=(2l_Bn_0)^{1/2}$
is the Debye wavenumber and $\omega=|\upsilon|/(l_B C)^{1/2}$ is a scaled $\upsilon$.
Here, $l_B=e^2/(\veps_0\veps(\phi_0)k_BT)$ is the Bjerrum length at $\phi_0$ and
$l=1/\kappa\xi$
is the ratio of the Debye length and the modified correlation length $\xi$:
$\xi=(C/\tau)^{1/2}$, where $\tau = \partial^2 f_m (\phi_0)/\partial\phi^2 -n_0(\Delta u
^++\Delta u ^-)^2/2$.

The solution of the linear equations with the $z\to-z$ symmetry is:
\begin{align}
\label{eq:solpb}
 \Psi(z)&=a_1 \cosh(q_1 z) -a_2 \cosh(q_2 z),\\
\label{eq:solcomp}
  \vphi(z)&=b_1 \cosh(q_1 z) -b_2 \cosh(q_2 z).
\end{align}
The wavenumbers $q_i$ obey
\begin{equation}
 \label{eq:solcond1}
  (q_i/\kappa)^4-\left(1+l^2-\omega^2\right)
( q_i/\kappa)^2+l^2=0.
\end{equation}
The amplitudes $a_i$ and $b_i$ are determined using the boundary conditions; in the
special case where the plates are electrically neutral they are
\begin{align}
\label{eq:ai}
 a_i=\frac{\Delta \gamma}{C}\frac{\kappa^2 \upsilon}{
q_i(q_2^2-q_1^2)\sinh(q_i
D/2)},\\
\label{eq:bi}
 b_i=\frac{\Delta \gamma}{C} \frac{
q_i^2-\kappa^2}{ q_i(q_2^2-q_1^2)\sinh(q_i
D/2)}
\end{align}

In the linear case it follows that
\begin{align}
\label{eq:potlin}
\frac{U}{k_BT}=S\times\frac{ (\Delta \gamma)^2}{2C}\biggl[&\Lambda_1\frac{
\coth(q_1 D/2)-1}{q_1} \nn \\ -&\Lambda_2 \frac{\coth(q_2 D/2)-1}{q_2}  \biggr]~,
\end{align}
where 
\begin{align}
\label{eq:Qdef}
\Lambda_i=\frac{q_i^2-\kappa^2}{q_2^2-q_1^2}.
\end{align}

One can see from \eqref{eq:potlin} that the interaction is $\propto(\Delta
\gamma)^2$ and that the interplay between the two terms in brackets
determines the nature of $U$. In the limit of
vanishing solvation asymmetry $\omega \rightarrow 0$, we have
from \eqref{eq:solcond1}: $q_1\rightarrow \kappa$ and $q_2\rightarrow\xi^{-1}$,
leading to $\Lambda_1\rightarrow0$ and $\Lambda_2\rightarrow$1. The resulting
potential is attractive, as expected when ion solvation is absent due to
critical adsorption \cite{Fisher1978}.

For non-vanishing $\omega$, we focus on the region above
$T_c$ for which $l>1+\omega$ and hence both $q_1$ and $q_2$ are positive real
numbers. For sufficiently small $\omega$ we find
\begin{equation}
\label{eq:qapx}
q_1\cong\kappa \sqrt{ 1+ \frac{\omega^2}{l^2-1}} \ , \ \ \ q_2\cong\xi^{-1}
\sqrt{ 1- \frac{\omega^2}{l^2-1}}.
\end{equation}
In this region, it is easy to show that $q_2>q_1>\kappa$ and thus
$\Lambda_1>0$ and $\Lambda_2>0$. Therefore, the
first term in brackets in \eqref{eq:potlin} is repulsive while the second is
attractive, leading to the existence of a maximum in the potential. The
location and magnitude of this repulsive barrier are found by solving $\partial
U/\partial D=0$. In the limit $D \gg q_1^{-1},q_2^{-1}$ we find for $D_{\rm max}$ and 
$U_{\rm max}$:
\begin{align}
\label{eq:analudmax}
D_{\rm max}=\frac{\log(\Lambda_2/\Lambda_1)}{q_2-q_1},\\
\frac{U_{\rm max}}{k_BT}= S\times\frac{ (\Delta
\gamma)^2}{C}\biggl[&\frac{
\Lambda_1}{q_1}\left(\frac{\Lambda_2}{\Lambda_1}\right)^{\frac{q_1}{q_2-q_1}} -
\frac{\Lambda_2}{q_2} 
\left(\frac{\Lambda_2}{\Lambda_1}\right)^{\frac{q_2}{q_2-q_1}}
\biggr].\nn
\end{align}
Whether $U_{\rm max}$ is significant depends on the ratio of amplitudes
$\Lambda_2/\Lambda_1$.

At large enough colloid separations $D$ the repulsive tail of the
interaction is $U/k_BT\simeq
S(\Delta\gamma)^2/(C/\Lambda_1)\exp(-q_1D)/q_1$. This
expression is analogous to the regular Debye-H\"{u}ckel result $U/k_BT\simeq
S(\sigma^2/\varepsilon_0\varepsilon)\exp(-\kappa D)/\kappa$
for charged  colloids \cite{colloids_book}. In our theory 
$q_1$ is a modified Debye wavenumber, $C/\Lambda_1$ is a property of the medium, and 
$\Delta \gamma$ plays the role of an effective
surface charge, reflecting the properties of the surface.

The comparison between the full
potential and the analytical approximation \eqref{eq:potlin} is shown in
\figref{fig_stab2}. 
For both $\upsilon =6$ ($\omega=0.64$) and $\upsilon =4$ ($\omega=0.43$), the linear
theory (dash-dot curves) agrees quite well with the numerical solution (solid
curves). Two circles in \figref{fig_stab2} show $D_{\rm max}$ and $U_{\rm max}$
evaluated using \eqref{eq:analudmax}. Our
analysis shows that $q_1$ and $\Lambda_1$ must be large enough for significant repulsion 
to appear. Hence, from \eqref{eq:qapx} and \eqref{eq:Qdef} we conclude that
$\omega$ should not be too small and $l$ not too large. The first requirement is satisfied
by choosing antagonistic salts while the second dictates the temperature window given
the salt concentration.

\begin{figure}[!tb]
\centering
\includegraphics[width=3.3in,clip]{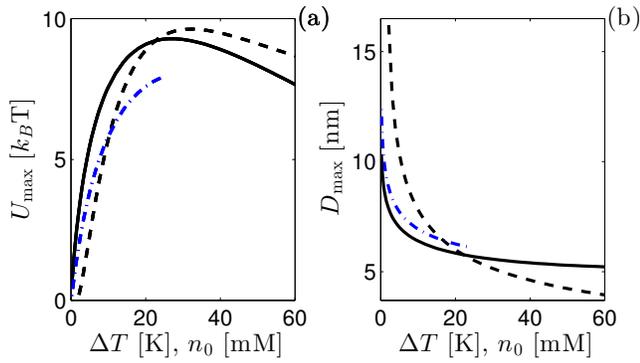}
\caption{(a) Barrier height and (b) location as a function of $\Delta T$ and
$n_0$ for $\upsilon=6$ and $\Delta\gamma=0.2/a^2$. Dashed curves are
numerical results for $\phi_0=0.5$,
$n_0=20$mM and varying $\Delta T$; a repulsive barrier appears at $\Delta
T\gtrapprox2$K. For
the solid curves $\phi_0=0.5$, $\Delta T=21.5$K and
$n_0$ varies. Dash-dot curves are \eqref{eq:analudmax} for varying $n_0$ plotted in the 
validity range given by $l>1+\omega$.}
\label{fig_stab3} 
\end{figure}

The two experimentally important quantities $U_{\rm max}$ and $D_{\rm max}$ are plotted in
\figref{fig_stab3} (a) and (b), respectively. For increasing $\Delta T$ (dashed curves) a
repulsive barrier first appears at $\Delta T \approx 2$K, and it has a maximal value. The
solid curves give results for varying $n_0$, showing
again a maximum in $U_{\rm max}$. Dash-dot lines are
$U_{\rm max}$ and $D_{\rm max}$ vs $n_0$ from \eqref{eq:analudmax} in the range $l>1+\omega$.
\figref{fig_stab3} (a) shows that stabilization can be achieved far 
above the critical temperature. Thus, in principle the theory applies to 
experiments with completely miscible mixtures, \textit{e.g}, water and alcohol, which in our 
theory is the $\chi \rightarrow 0$ limit of athermal mixtures.

The behavior of $U_{\rm max}$ is determined by the
interplay between the attractive adsorption-related part of interaction and the repulsive
solvation-related part. An increase in $\Delta T$ or $n_0$ increases $\xi^{-1}$ or
$\kappa$, respectively. The result in both cases is an increase in the wavenumbers $q_i$
and therefore a decrease in $D_{\rm max}$, see \figref{fig_stab3} (b). Furthermore, the
relative magnitudes of the wavenumbers $q_i$ and the amplitudes $\Lambda_i$ change in a
non
trivial manner due to the coupling of the attractive and repulsive contributions, given by
the parameter $l$ in the linear theory.

\begin{figure}[!tb]
\centering
\includegraphics[width=3.3in,clip]{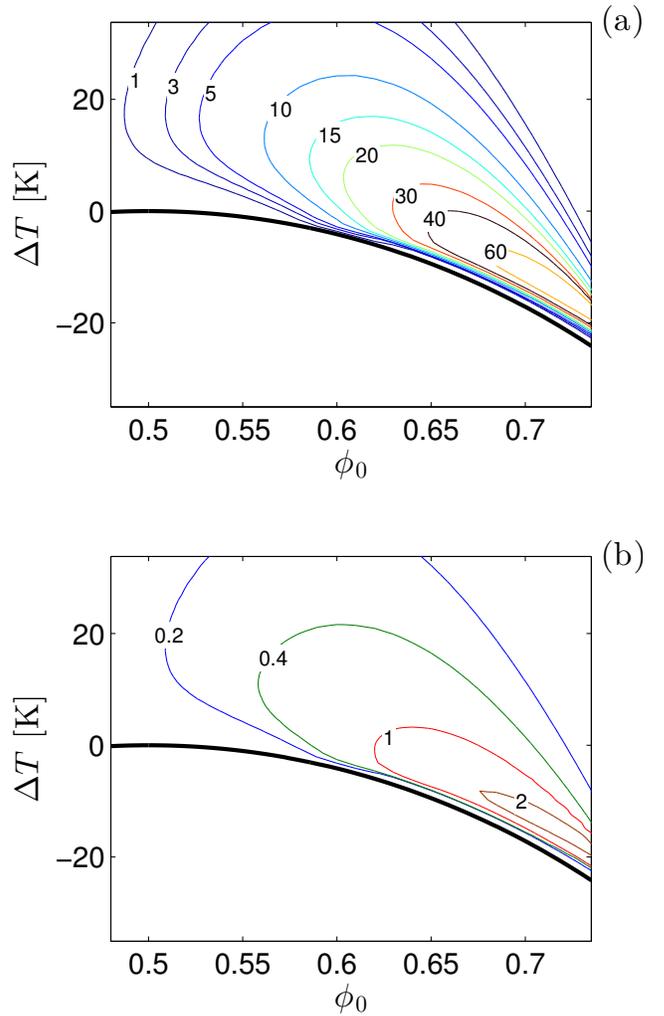}
\caption{The maximum of the interaction potential, $U_{T,{\rm max}}(\phi_0,T)$, in the
$\phi_0$--$T$ plane
including vdW attraction and using Derjaguin's approximation for a colloid radius $R=1\mu$m. (a) Contours
of $U_{T,{\rm max}}$ for a Hamaker constant
$A_H=0.2\times10^{-20}$J. (b ) The contour lines corresponding to  
$U_{T,{\rm max}}=3k_BT$ for different Hamaker constants. The values of Hamaker's 
constant are indicated by the labels in units of $10^{-20}$J. Here $\upsilon=6$, 
$n_0=20$mM and $\Delta\gamma=0.2/a^2$.}
\label{fig_stab4} 
\end{figure}

In the spirit of the DLVO theory, the more realistic case of spherical colloids of radius
$R$ is evaluated by applying Derjaguin's approximation to the potential $U$ and adding the
vdW interaction between the spheres: $U_T=\pi R \int_D^\infty U(D'){ \rm d} D'-A_HR/(12
D)$, where $A_H$ is the Hamaker constant. Contours of the maximum of
$U_T$, $U_{T,{\rm max}}$, are shown in \figref{fig_stab4} (a) in the
$\phi_0$--$T$ plane. Notice that $U_{T,{\rm max}}$ increases significantly for
water-rich compositions ($\phi_0>0.5$). The reason for this is
twofold: (i) the adsorption force is weaker at off-critical compositions and (ii) relative
to the bulk composition, the water-poor layer on the surface is more attractive for
hydrophobic ions. For hydrophilic colloids ($\Delta\gamma<0$) $U_{T,{\rm max}}$
would be larger at compositions $\phi_0<0.5$. While the absolute values of
$U_{T,{\rm max}}$ do not depend on the sign of $\Delta\gamma$, the
hydrophobicity or hydrophilicity of the colloids determines the ideal working
region in the $\phi_0$--$T$ plane. The contour lines of $U_{T,{\rm max}}=3k_BT$
are shown in \figref{fig_stab4} (b) for different values of the Hamaker constant. 
The area enclosed by this contour defines approximately the working conditions for a
stable dispersion. Indeed, even for large values of the Hamaker constant 
a stable region exists for water-rich compositions and closer to the binodal curve. 
Hence, one could add only a small amount of 
co-solvent to an unstable dispersion of charge-free colloids to obtain a stable dispersion.

\begin{figure}[!tb]
\centering
\includegraphics[width=3.3in,clip]{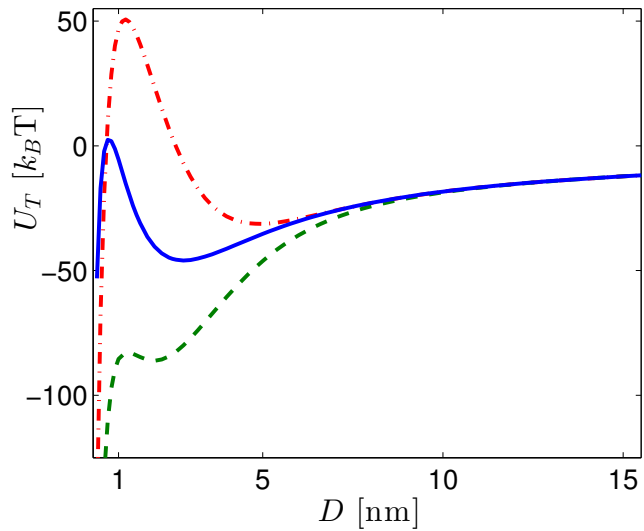}
\caption{$U_T(D)$ for two charged colloids with a surface charge
density $\sigma=0.1$nm$^{-2}$. When $\Delta \gamma =0$ and $\Delta u^\pm=0$
(solid
curve) the potential is marginally stable. For a hydrophobic surface, $\Delta
\gamma=0.2/a^2$, the potential is unstable for a hydrophilic salt ($\Delta u^\pm=6$,
dashed
curve) while it has a large barrier for an antagonistic salt ($\Delta
u^+=-\Delta u^-=6$, 
dash-dot curve). Here $n_0=20$mM, $\phi_0=0.5$,
$\Delta T=21.5$K, $R=1\mu$m and $A_H=1\times10^{-20}$J.}
\label{fig_stab5} 
\end{figure}

The linear solvation model adopted by us for its simplicity is a first-order 
approximation, and the study of more complex and realistic solvation models is an active 
area of research \cite{bier2012}. Nonetheless, the large repulsive barriers we predict are 
not restricted to a linear solvation model \cite{comment1}. In addition, a significant 
barrier can be obtained also for weakly antagonistic salts ($\upsilon\simeq1$) if the 
plates are made more hydrophobic or hydrophilic (\eqref{eq:potlin}), or if the bulk 
composition is changed (\figref{fig_stab4} (a)). 

In \figref{fig_stab5} we show that addition of antagonistic salts can enhance the
stability of \emph{charged} colloids as well. When the colloids' surface and the ions are
indifferent to the solvents, $\Delta \gamma =0$ and $\Delta u^\pm=0$, as in the regular
Poisson-Boltzmann 
theory, the effect of the mixture on the interaction is via the dependence of the
dielectric constant on $\phi$. The result for $U_T(D)$, shown by the solid curve in
\figref{fig_stab5}, is a marginally stable potential. For hydrophobic colloids
($\Delta \gamma >0$), $U_T$ becomes attractive if the salt is hydrophilic
(dashed curve), indicating a destabilization of the suspension. However, for
hydrophobic colloids
and an antagonistic salt (dash-dot curve) the repulsive barrier increases
significantly,
indicating a stabilization of the suspension. Here, the hydrophobic and positively charged
colloids draw a larger amount of hydrophobic anions towards their surface, leading to an
increase in the repulsive osmotic pressure of ions.

In conclusion, the theory predicts that significant potential barriers exist in
a wide temperature and composition range and shows that neutral and charged colloids can 
be effectively suspended in a binary mixture by addition of antagonistic salts. It
is worth noting that the specific adsorption of ions to the surface not
discussed here will typically enhance the stabilization. The ion affinity to the
wetting liquid will usually come hand in hand with a similar surface
affinity thus enhancing the electrostatic repulsion. 

The mechanism we describe is of potential use in numerous colloidal systems where 
currently only one solvent is employed and it is advantageous over existing methods in 
cases where the modification of the colloid surface chemistry is undesired. The most important 
requirement to achieve a stable suspension is to chose a salt in which the ions' 
solvation asymmetry in the mixture is large enough.  We expect our results to be most 
beneficial for dispersing charge-free particles. For example, there have been large 
efforts recently in ``transparent and conducting'' electrodes for solar cell applications 
\cite{regev2013}. In these works graphite is typically sonicated to yield graphene 
sheets, and these sheets are dispersed using surfactants. Using evaporation or slow 
sedimentation these sheets assemble as a thin and conducting layer on top of a 
transparent substrate. In those works surfactants stay between graphene sheets 
and reduce the conductivity immensely. In this and other cases, using salts for the 
dispersion instead of surfactants could increase markedly the conductivity of the film.

\noindent
{\bf Acknowledgments} This work was supported by the Israel Science Foundation under 
grant No. 11/10, the COST European program MP1106 ``Smart and green interfaces
 - from single bubbles and drops to industrial, environmental and biomedical 
applications'',  and the European Research Council ``Starting Grant'' No. 
259205.


\begin{thebibliography}{10}
\bibitem{comment1}
{\rm We verified that our conclusions are not restricted to the choice of a
  bilinear solvation energy. Using data from Refs.
  \cite{marcus_cation,marcus_anion} we have fitted, using polynomials, the
  experimental solvation energy for antagonistic salts. For this more complex
  model, our results remain qualitatively the same and quantitatively similar.}
\end{thebibliography}

\end{document}